\documentclass[]{aa}
\usepackage{graphicx}
\usepackage{natbib}
\usepackage{psfig}
\usepackage{txfonts}
\newcommand\eg{{\it e.g.} }
\newcommand\ie{{\it i.e.} }
\newcommand\etal{et~al.}
\newcommand\HI{\ion{H}{I}}

\newcommand\Lya{Ly$\alpha$}

\newcommand\HeII{\ion{He}{II}~$\lambda$~1640}

\newcommand\kms{\ifmmode {\rm\,km\,s^{-1}}\else${\rm\,km\,s^{-1}}$\fi}
\newcommand\Msun{M$_\odot$}
\newcommand\Lsun{L$_\odot$}
\font\aipsfont = cmsy8 scaled\magstep1

\newcommand\aips {{\aipsfont AIPS}}
\def\spose#1{\hbox to 0pt{#1\hss}}
\newcommand\simlt{\mathrel{\spose{\lower 3pt\hbox{$\mathchar"218$}}
     \raise 2.0pt\hbox{$\mathchar"13C$}}}
\newcommand\simgt{\mathrel{\spose{\lower 3pt\hbox{$\mathchar"218$}}
     \raise 2.0pt\hbox{$\mathchar"13E$}}}
\begin{document}
\title{Detection of Two Massive CO Systems in 4C~41.17 at z = 3.8}

\author{C. De Breuck \inst{1}
        \and
        D. Downes\inst{2}
        \and
        R. Neri\inst{2}
        \and
        W. van Breugel\inst{3}
        \and
        M. Reuland\inst{3,4}
        \and
	A. Omont\inst{5}
	\and
        R. Ivison\inst{6,7}
          }


\institute{European Southern Observatory, Karl Schwarzschild Stra\ss e 2, D-85748 Garching, Germany
\and Institut de Radioastronomie Millim\'etrique, Domaine Universitaire, F-38406 St. Martin-d'H\`eres, France
\and IGPP/LLNL, L-413, 7000 East Ave, Livermore, CA 94550, USA
\and Sterrewacht Leiden, Postbus 9513, NL-2300 RA Leiden, The Netherlands
\and 
Institut d'Astrophysique de Paris, CNRS \& Universit\'e Paris 6, 98bis Boulevard Arago, F-75014 Paris, France
\and Astronomy Technology Centre, Royal Observatory, Blackford Hill, Edinburgh EH9 3HJ, United Kingdom
\and Institute for Astronomy, University of Edinburgh, Royal Observatory, Blackford Hill, Edinburgh EH9 3HJ, United Kingdom
}

\date{Received 2004 October 20; accepted 2004 November 26}

\abstract{We have detected CO(4$-$3) in the $z$=3.8 radio galaxy
4C~41.17 with the IRAM Interferometer. The CO is in two massive
($M_{\rm dyn} \sim 6 \times 10^{10}$M$_{\odot}$) systems separated
by 1$\farcs$8 (13~kpc), and by 400\kms\ in velocity, which coincide
with two different dark lanes in a deep \Lya\ image. One CO component
coincides with the cm-radio core of the radio galaxy, and its redshift
is close to that of the \HeII\ AGN line.  The second CO component is
near the base of a cone-shaped region southwest of the nucleus, which
resembles the emission-line cones seen in nearby AGN and starburst
galaxies.  The characteristics of the CO sources and their mm/submm
dust continuum are similar to those found in ultraluminous IR galaxies
and in some high-$z$ radio galaxies and quasars.  The fact that
4C~41.17 contains two CO systems is further evidence for the role of
mergers in the evolution of galaxies at high redshift.

  }

\maketitle
%

\section{Introduction}
Because of its luminosity and large angular extent, 4C~41.17
has become the most studied high redshift radio galaxy (HzRG). {\it
HST} images show that its host galaxy contains several star-forming
components, including (i) a linear radio-aligned feature with
spectroscopic characteristics of a young stellar population
\citep{dey97}, (ii) a clumpy system, separate from the radio source,
and (iii) low surface brightness UV emission extending over 70~kpc
\citep{wvb99}. The entire system is embedded in a giant \Lya\ halo
\citep{reu03} and diffuse soft X-ray emission \citep{sch03}. In spite
of the detection of its dust emission \citep{dun94,chi94}, searches
for molecular gas have been unsuccessful up to now
\citep{ivi96,eva96,bar96,sco96}. In this letter, we report the first
detection of two massive CO components near the centre of this large
forming galaxy\footnote{We adopt H$_0$=71\,km\,s$^{-1}$\,Mpc$^{-1}$,
$\Omega_{\rm M}$=0.27 and $\Omega_{\Lambda}$=0.73. At $z$=3.8, the
luminosity distance $D_L$=34.4\,Gpc, and 1\arcsec\ corresponds to
7.2\,kpc.}.

\section{Observations and data reduction}
We observed 4C~41.17 at 3.1 and 1.2\,mm with the IRAM Plateau de Bure Interferometer
in its compact $D$ configuration between 1999 August and 2003 August.
The spectral correlator covered a bandwidth of 580~MHz at each
frequency. At 3.1\,mm, the 6-antenna equivalent on-source observing
time was 29\,h, and the beam was $7\farcs0 \times 4\farcs7$ at PA
107\degr.  The central frequency was initially tuned to 96.090~GHz to
observe CO(4$-$3) ($\nu_{\rm rest}=461.040$~GHz, but later re-centred
to 96.250~GHz to better cover the detected line.  After phase and
amplitude calibration, we merged the data at different centre
frequencies into a data-set of 36 channels of 20~MHz each. As a
result, the outer 7 channels on each end have $\sigma$$\approx$0.9\,mJy compared to $\sigma$$\approx$0.6\,mJy in the central ones.  In the final data cube,
the zero point of the velocity scale is 96.093~GHz, corresponding to
$z=3.79786$, the redshift of the \HeII\ line \citep{dey97}. We made
naturally weighted maps using the \aips\ task {\tt IMAGR}, with
CLEANing applied only to the velocity-integrated maps.
\begin{figure}[ht]
\psfig{file=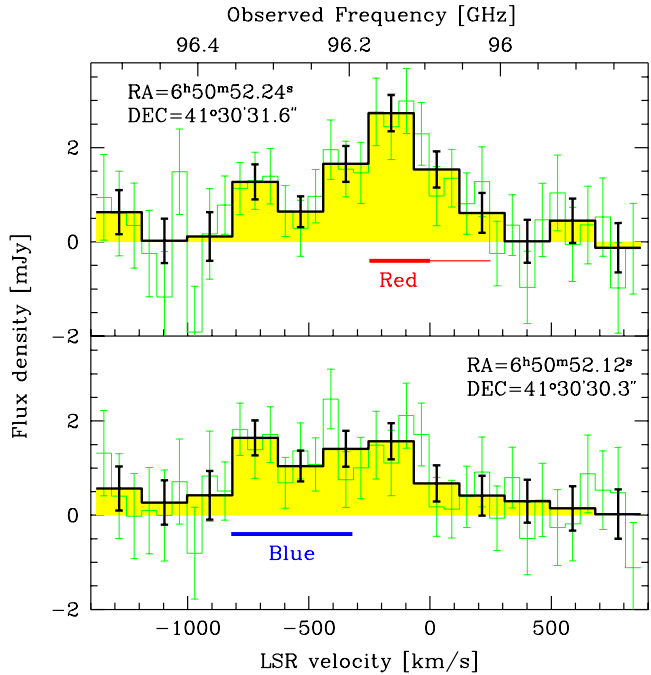,width=9cm}
\caption{CO~(4$-$3) spectra at the positions of the red (top panel)
and blue components (bottom). The thin green line shows the data in
62\,\kms\ channels, and the thick black line in 184\,\kms\
channels. Error bars are 1$\sigma$, and velocities are relative to
96.093~GHz ($z_{\rm LSR} =$ 3.79786). The red and blue bars indicate the linewidths; the thick part of the red bar marks channels used in Fig.~3, left. Note that these two positions are less than a beam size away, and are therefore not fully independent.}
\label{cospec}
\end{figure}

We observed simultaneously at 1.2\,mm (241.453\,GHz DSB) to study the
dust continuum. We used only 1.2~mm data taken with a precipitable
water vapour content $<3$~mm, which gave a usable on-source observing
time of 7.7~h, and an rms noise of 0.8~mJy/beam. The 1.2\,mm beam
was $2\farcs9 \times 1\farcs5$ at PA 100\degr, but we convolved the image
with a 3\farcs0$\times$3\farcs0 Gaussian.

In 2004 March, we also observed the 4C~41.17 field at 1.2\,mm ($\sim$250~GHz)
with the 117-element MPIfR Millimeter Bolometer array
\citep[MAMBO-2;][]{kre98} at the IRAM 30m telescope. The beam FWHM is
10\farcs7 with an array size of 4\arcmin.  We made eight 
on-the-fly maps, with 41 subscans of 40~s each,
while chopping the secondary mirror in azimuth at 2~Hz by 39, 42, or
45\arcsec. We reduced the data using MOPSIC \citep{zyl98}. The
map covers 3\arcmin$\times$3\arcmin\ with an rms noise of
0.8 mJy.

\begin{table}
\caption{Observed and derived parameters for CO(4$-$3) in 4C~41.17.}
\begin{scriptsize}
\begin{tabular}{lrrr}
Parameter & Red component & Blue component & Total \\
\hline
RA(J2000)$^a$ (peak) & 6$^{\rm h}$50$^{\rm m}$52\fs24 & 6$^{\rm h}$50$^{\rm m}$52\fs12 & 6$^{\rm h}$50$^{\rm m}$52\fs17 \\
DEC(J2000)$^a$ (peak) & 41\degr30\arcmin31\farcs6 & 41\degr30\arcmin30\farcs3 & 41\degr30\arcmin30\farcs9 \\
$\int S_{\rm CO}\,dV$ [Jy \kms] & $1.2\pm 0.15$ & $0.6\pm 0.15$ & $1.8\pm 0.2$\\
Central velocity $^{\mathrm{b}}$ [\kms]    &$-130\pm 50$ &$-550\pm 100$ & $-$285$\pm$100 \\
Velocity width [\kms] &$500\pm 100$ &$500\pm 150$ & 1000$\pm$150 \\
$L^{\prime}_{\rm CO}$(4$-$3) [K\kms pc$^2$] & 4.4$\times 10^{10}$ & 2.2$\times 10^{10}$ & 6.7$\times 10^{10}$ \\
M(H$_2$) [M$_{\odot}$] & 3.6$\times 10^{10}$ & 1.8$\times 10^{10}$ & 5.4$\times 10^{10}$\\
\hline
\end{tabular}
\end{scriptsize}
\begin{list}{}{}
  \item[$^{\mathrm{a}}$]
Positional uncertainty 0\fs03 in RA and 0\farcs3 in DEC
  \item[$^{\mathrm{b}}$]
Relative to 96.093\,GHz ($z$ =3.79786).
\end{list}
\label{obsparameters}
\end{table}
%

\begin{figure}[ht]
\psfig{file=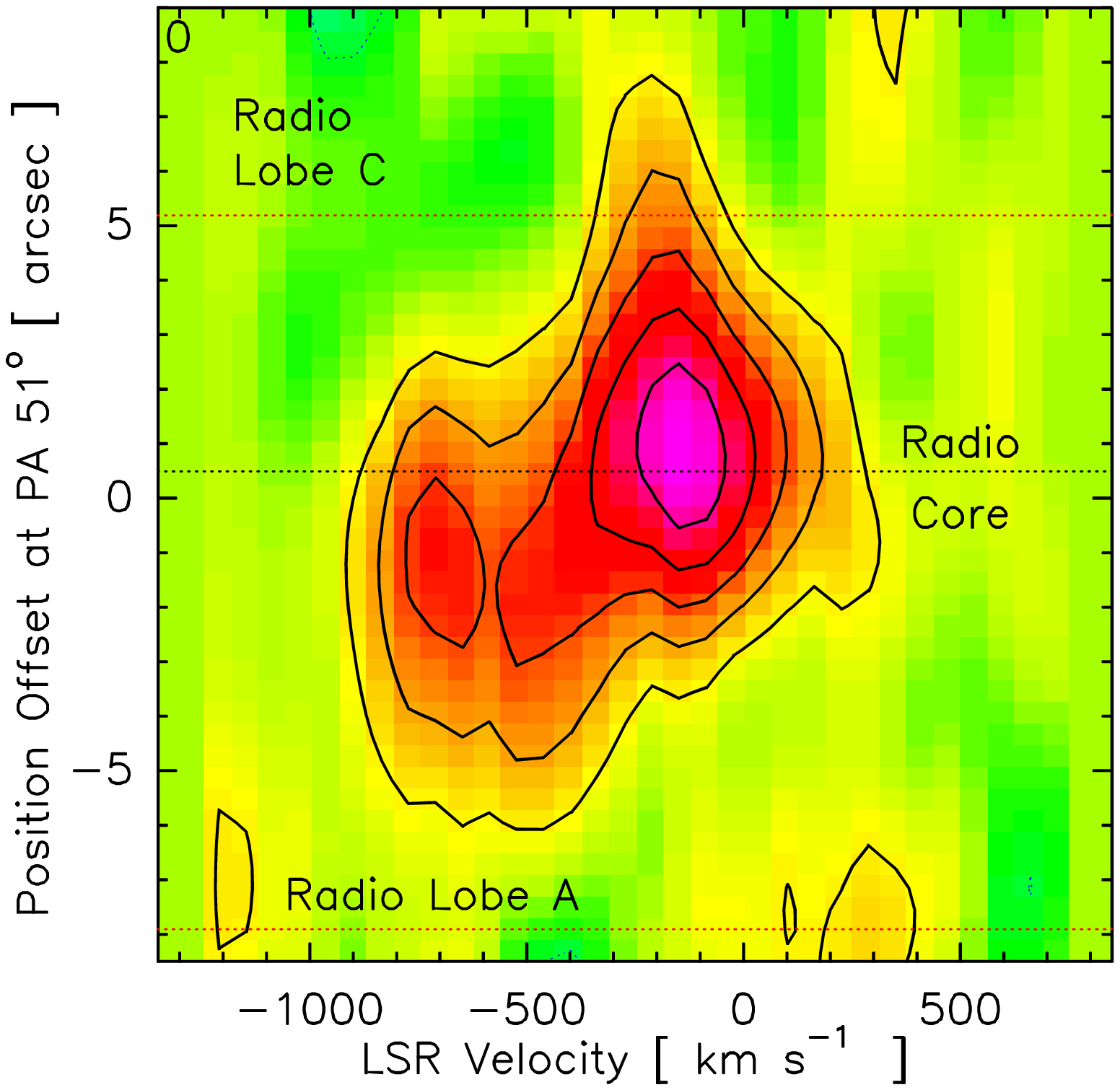,width=8.5cm,angle=-90}
\caption{CO~(4$-$3) position-velocity slice at PA=51\degr\ (\ie\ along the radio
axis). Horizontal lines mark the radio core and outer lobes. 
The position zero is 6$^{\rm h}$50$^{\rm m}$52\fs17 
$41\degr30\arcmin30\farcs9$, the beam along this cut is 
6\arcsec, and the contour steps are 0.4\,mJy, with the first contour at 0.4\,mJy.}
\label{coslice}
\end{figure}

\section{Results}
Figure~\ref{cospec} shows CO(4$-$3) spectra at two positions.  No line
emission is detected in the range $-$1550 to $-$1000\,\kms\ and +250
to +800\,\kms\ to 0.75\,mJy (3$\sigma$). In these outer channels, we
find a marginal continuum emission at $\sim$0.3\,mJy, which is
consistent with the 11\,mJy at 850\,$\mu$m \citep{ivi00} and 3.8\,mJy
at 1.2\,mm (our MAMBO map) extrapolated to 3.1\,mm
($\sim$0.2\,mJy), and the non-thermal contribution of $\simlt$0.2\,mJy
(over the 13\arcsec\ source), extrapolated from 6.4\,mJy at 2\,cm
(Chambers \etal, 1990)\nocite{cha90} and 2.7\,mJy at 1.2\,cm \citep{ivi96}.  Because of these
low values (below the first contour in Fig.\,\ref{coslice}), we do not
correct the CO fluxes for the continuum.

\begin{figure*}[ht]
\psfig{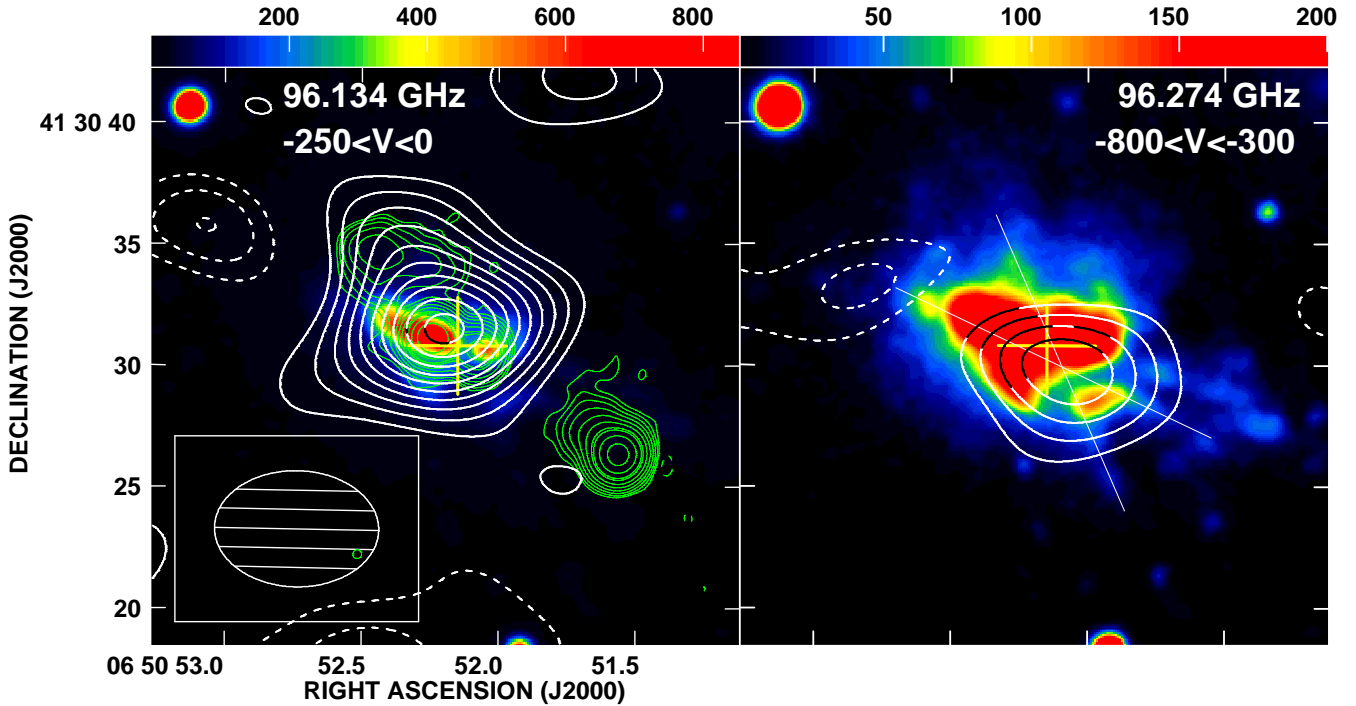}
\caption{Velocity-averaged CO maps (thick/white+black contours) of
the red (left) and blue (right) components.  CO contours are $-$3,
$-$2, 2, 3, ..., 10$\sigma$, with $\sigma$=0.2 mJy~beam$^{-1}$. The
lower left inset shows the CO beam.  The two CO maps are superposed on the
\Lya\ image, each with its own colour scheme to show the two dark lanes in the Ly$\alpha$.
The thin green contours
({\it left panel}) show the 1.4~GHz radio map (Carilli et al.\ 1994), 
and the
yellow cross indicates the radio core. The thin straight lines ({\it
right panel}) indicate a possible AGN or starburst emission-line
cone.}
\label{LyaCOradio} \end{figure*} %

The position-velocity slice (Fig.~\ref{coslice}) shows that the CO
emission has two components: (i) a 'red' component at $\sim -$130\kms\
relative to $z$=3.79786, and (ii) a 'blue' component at
$-$550\kms.  Figure~\ref{LyaCOradio} shows the integrated red and blue
components\footnote{To increase S/N, the red component in Fig.~\ref{LyaCOradio} covers only the brightest 250\kms, indicated by the thick bar in Fig.~\ref{cospec} (top).}, separated by
1\farcs8 (13\,kpc projected).  Table~\ref{obsparameters} lists the
observed parameters, the line luminosity $L^{\prime}_{\rm
CO}$(4$-$3), and the molecular gas mass M(H$_2$), calculated assuming
a constant brightness temperature from CO(4$-$3) to CO(1$-$0) and a
conversion factor $X_{\rm CO}$=0.8~M$_{\odot}$(K\kms pc$^2$)$^{-1}$ 
derived for local ultraluminous infrared galaxies \citep{dow98}.

\begin{figure}[ht]
\psfig{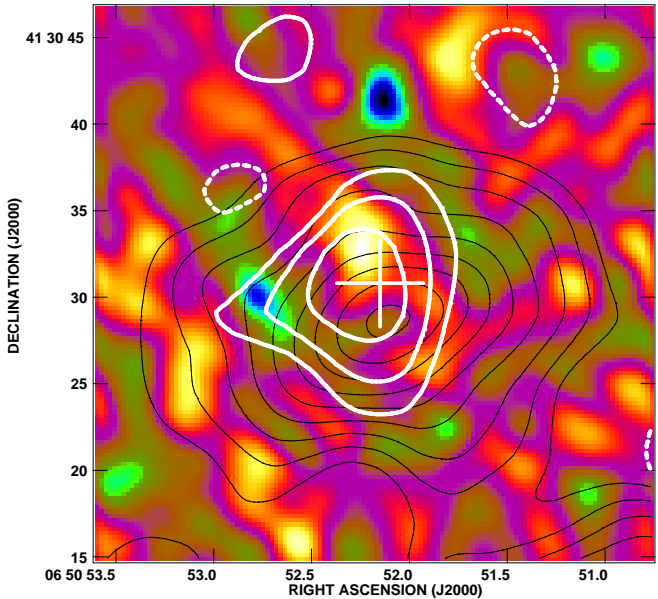}
\caption{
Thermal dust emission in 4C~41.17. The colourscale shows the 1.2\,mm
PdBI map convolved with a 3\farcs0 Gaussian (not corrected for the
20$''$ primary beam).  Superposed are the 850\,$\mu$m SCUBA map (thin
black contours; beam 14$''$) and our 1.2\,mm MAMBO map (thick white
contours; beam 11$''$).  Contours are -2, 2, 3, 4, 5, 6, 7, 8, 9 and
10$\sigma$, with $\sigma$=1.2 and 0.8 mJy~beam$^{-1}$ for the SCUBA
and MAMBO maps, respectively. The white cross indicates the cm-radio
core. }
\label{dustimages}
\end{figure}

Figure~\ref{dustimages} shows three maps of the dust continuum in
4C~41.17. Unlike the solid 10$\sigma$ and 5$\sigma$ detections in
the 850\,$\mu$m SCUBA and 1.2\,mm MAMBO maps, our 1.2\,mm
interferometer map shows a 4.3$\sigma$ peak at 06$^{\rm h}$50$^{\rm
m}$52\fs24, +41\degr30\arcmin31\farcs9 (J2000; \ie at the red CO
component) only after convolution with a 3\farcs0 Gaussian. This
PdBI map yields a flux $S_{\rm 1.2\,mm}=3.0\pm 0.7$~mJy,
consistent with the $S_{\rm 1.2\,mm}=3.8\pm 0.6$~mJy from the MAMBO
map. The S/N in our 1.2\,mm maps is insufficient to constrain the
spatial extent of the thermal dust emission reported by \citet{ivi00}
and \citet{ste03}.

Both CO components are gas-rich systems with $\rm M_{\rm H_2}\sim
3\times 10^{10}$\,\Msun\ (Table~\ref{obsparameters}). Their projected
separation of 13~kpc and apparent relative velocity of $\sim$400\kms\
imply a combined dynamical mass $R\Delta V^2/G$ of $\simgt 6 \times
10^{10}$M$_{\odot}$, an order of magnitude smaller than the typical
baryonic mass of HzRGs integrated over 64\,kpc \citep{roc04}.

We associate the red CO component with the AGN because: (i) it agrees
within the uncertainties, with the cm-radio core position \citep{car94} and
the hard X-ray point source \citep{sch03}, (ii) it coincides with the
central dark lane seen in the \Lya\ image (Fig.~\ref{LyaCOradio}, {\it
left,} near the cross) and the HST $R-$band image \citep{bic00}, and
(iii) it has the same velocity offset as the low-ionization
interstellar absorption lines in the deep Keck spectrum of
\citet{dey97}. The CO redshift of 3.7958$\pm$0.0008 is thus likely to
be the real systemic redshift of the host galaxy of 4C~41.17.

The blue CO component peaks at the position of a second apparent dark
lane in the \Lya\ image (Fig.~\ref{LyaCOradio}, {\it right}), which
suggests that the \Lya\ may be absorbed by dust of another kinematic
system, associated with this CO. This apparent \Lya\ absorption lane
is in a region which resembles the emission-line cones seen, although
on smaller scales (a few kpc), in some nearby AGN \citep[\eg\
Cygnus~A; ][]{can03} as well as in starburst galaxies \citep[\eg][]{vei02}.

\section{Discussion and Conclusions}

What is the relation between the molecular CO-gas, the dust, the \Lya\
halo, and the massive forming galaxy with its radio-loud AGN?
Our CO velocity profile (Fig.~\ref{coslice}) shows a remarkable
similarity to the \Lya\ velocity profile of \citet{dey99}. Both the CO
and \Lya\ are split into two components, separated by a projected
distance of $\sim$13~kpc, and by $\sim$400\,\kms. This suggests that
the \Lya\ emission may also come from two separate components
(rather than a single component split by an associated \HI\
absorber). However, \Lya\ traces much less dense gas
\citep[$n_{\rm H} \sim$ 17--150 cm$^{-3}$;][]{vil03} than CO,
which must have a density of 10$^3$ to 10$^4$cm$^{-3}$ to have enough
CO-line opacity to give a typical brightness temperature of order 30\,K. 
Hence, the CO and \Lya\ may originate from the same gas-rich systems,
but they do not necessarily trace the total extent of these
regions. This is obvious from the 200\,kpc spatial extent of the \Lya,
while the CO emission is unresolved with our 6\arcsec\ (43\,kpc)
beam. If fact, we can put even stronger constraints on the size of the
CO-line sources, using some basic assumptions.

The FIR dust luminosity of 4C41.17 is very high, $L_{\rm FIR} \approx
1.5\times 10^{13}$\,\Lsun , a value typical of an ultraluminous
starburst.  The FIR fluxes, including our new continuum data points at
3.1 and 1.2\,mm, imply a dust temperature of $54\pm 10$\,K, in
agreement with earlier values \citep{ben99,sch03}.  Although the
mm/sub-mm continuum is optically thin, the CO lines are not, so the
observed brightness temperature of the CO will be about the same as
the gas temperature.  The existing interferometer CO maps of ULIRGs,
the only nearby objects with comparable FIR luminosity, show that the
brightness temperatures of the low-J CO lines are comparable with the
FIR dust temperature \citep{dow98}.  Indeed, for high-$z$ CO detections even
to be possible, the gas must have a significant brightness
temperature, typically 30 to 50\,K, over several hundred pc.
For 4C41.17, this means that if the CO(4$-$3) brightness temperature
is $\sim$30\,K, then the observed CO(4$-$3) luminosity (Table~1)
implies a CO source diameter $d$=$(4\,L^\prime_{\rm CO}/(\pi\,T_b\,
\Delta V))^{0.5}$ of 1.4 to 1.8\,kpc, or 0\farcs2 to 0\farcs25.

This size and H$_2$ mass imply a hydrogen column density of order
$10^{24}$\,cm$^{-2}$, which is consistent with the observed mm-FIR
dust spectrum becoming opaque near restframe 100\,$\mu$m. Applying the
Stefan-Boltzmann formula with the derived CO source diameter
$d$$\sim$1.6\,kpc and dust temperature then yields $L_{\rm FIR} =
\pi\sigma\,d^2T^4$$\approx$10$^{13}$\,\Lsun, similar to the observed
value. The CO source size is also roughly consistent with the total
mass and typical density of the CO gas (10$^3$ to 10$^4$cm$^{-3}$),
which imply source sizes of order $\sim$1~kpc, depending on geometry.
Note that these are maximum sizes of the CO.  If one assumes the CO is
in more than two sources, then each component will have a smaller
size. We also note that our observations are insensitive to more widely
distributed, cooler CO.

The only known place where such high gas densities over such
dimensions are found are the circumnuclear disks observed in ULIRGs
and some quasars.  
The `red' CO source in 4C41.17 thus finds a natural interpretation as
a circumnuclear starforming disk around the radio-loud AGN.  
The fact that there appears to be a second such CO source, 13\,kpc
away, separated by $\sim$400\,\kms\ in velocity, that also contains
about the same mass of molecular gas, suggests the AGN and possibly
starburst activity may have been triggered by the interaction of the
two objects. The absence of UV/optical continuum signatures of
starburst activity at the position of the blue CO component remains 
surprising. A possible explanation could be that this starburst has not yet 
reached its peak UV/optical emission \citep{haas03}.

To summarize: our observations indicate that the 4C~41.17 system
contains two massive CO components, each of which may be associated
with an obscured black hole. This is remarkably similar to the two CO
systems in 4C~60.07 \citep{pap00,gre04}. There are three other HzRGs,
6C~1909+72, B3~J2330+3927 and TN~J0120+1320 with detected CO
\citep{pap00,deb03a,deb03b}.  It will be of interest to determine if
they are also double sources in CO, which would further indicate the
role of mergers in triggering AGN activity in the most massive galaxies at
high redshift.

\begin{acknowledgements}
IRAM is supported by INSU/CNRS (France), MPG (Germany) and IGN (Spain). 
The work by WvB and MR was performed at IGPP/LLNL under the auspices of the
U.S. Department of Energy, National Nuclear Security Administration by
the University of California, Lawrence Livermore National Laboratory
under contract No. W-7405-Eng-48. This work was carried out in the
context of EARA, the European Association for Research in Astronomy.
\end{acknowledgements}


\begin{thebibliography}{}
\bibitem[Barvainis \& Antonucci(1996)]{bar96} Barvainis, R. \& Antonucci, R. 1996, \pasp, 108, 187
\bibitem[Benford \etal(1999)]{ben99} Benford, D., Cox, P., Omont, A., Phillips, T., \& McMahon, R. 1999, \apj, 518, L65
\bibitem[Bicknell \etal(2000)]{bic00} Bicknell, G., Sutherland, R., van Breugel, W., Dopita, M., Dey, A., \& Miley, G. 2000, \apj, 540, 678
\bibitem[Canalizo \etal(2003)]{can03} Canalizo, G., Max, C., Whysong, D., Antonucci, R., \& Dahm, S. 2003, \apj, 597, 823
\bibitem[Carilli, Owen, \& Harris(1994)]{car94} Carilli, C., Owen, F., \& Harris, D. 1994, \aj, 107, 480 
\bibitem[Chambers, Miley, \& van Breugel(1990)]{cha90} Chambers, K., Miley, G., \& van Breugel, W. 1990, \apj, 363, 21 
\bibitem[Chini \& Kr\"ugel(1994)]{chi94} Chini, R. \& Kr\"ugel, E. 1994, \aap, 288, L33 
\bibitem[De Breuck \etal(2003a)]{deb03a} De Breuck, C., \etal\ 2003a, \aap, 401, 911
\bibitem[De Breuck \etal(2003b)]{deb03b} De Breuck, C., \etal\ 2003b, New Astron.\ Rev., 47, 285 
\bibitem[Dey \etal(1997)]{dey97} Dey, A., van Breugel, W., Vacca, W., \& Antonucci, R. 1997, \apj, 490, 698 
\bibitem[Dey(1999)]{dey99} Dey, A. 1999, ASP Conf.~Ser.~193: The Hy-Redshift Universe: Galaxy Formation and Evolution at High Redshift, ed.\ A.J.\ Bunker \& W.J.M. van Breugel, (ASP: San Francisco) 34 
\bibitem[Downes \& Solomon(1998)]{dow98} Downes, D., \& Solomon, P. 1998, \apj, 507, 615 
\bibitem[Dunlop \etal(1994)]{dun94} Dunlop, J., Hughes, D., Rawlings, S., Eales, S., \& Ward, M. 1994, \nat, 370, 347 
\bibitem[Evans \etal(1996)]{eva96} Evans, A., \etal\ 1996, \apj, 457, 658 
\bibitem[Greve, Ivison, \& Papadopoulos(2004)]{gre04} Greve, T., Ivison, R., \& Papadopoulos, P. 2004, \aap, 419, 99 
\bibitem[Haas \etal(2003)]{haas03} Haas, M., \etal\ 2003, \aap, 402, 87 
\bibitem[Ivison \etal(1996)]{ivi96} Ivison, R., Papadopoulos, P., Seaquist, E., \& Eales, S. 1996, \mnras, 278, 669 
\bibitem[Ivison \etal(2000)]{ivi00} Ivison, R., Dunlop, J., Smail, I.\ et al.\ 2000, \apj, 542, 27
\bibitem[Kreysa \etal(1998)]{kre98} Kreysa, E., \etal\ 1998, \procspie, 3357, 319 
\bibitem[Papadopoulos \etal(2000)]{pap00} Papadopoulos, P., \etal\ 2000, \apj, 528, 626 
\bibitem[Reuland \etal(2003)]{reu03} Reuland, M., \etal\ 2003, \apj, 592, 755 
\bibitem[Rocca-Volmerange \etal(2004)]{roc04} Rocca-Volmerange, B., Le Borgne, D., De Breuck, C., Fioc, M., \& Moy, E. 2004, \aap, 415, 931 
\bibitem[Scharf \etal(2003)]{sch03} Scharf, C., Smail, I., Ivison, R., Bower, R., van Breugel, W., \& Reuland, M. 2003, \apj, 596, 105
\bibitem[Scoville, Yun, \& Bryant(1996)]{sco96} Scoville, N., Yun, M., \& Bryant, P. 1996, ASSL Vol.~206: Cold Gas at High Redshift, ed.\ M.N.\ Bremer et al.\ (Kluwer: Dordrecht), 25 
\bibitem[Stevens \etal(2003)]{ste03} Stevens, J., \etal\ 2003, \nat, 425, 264 
\bibitem[van Breugel \etal(1999)]{wvb99} van Breugel, W., \etal\ 1999, 
in The Most Distant Radio Galaxies, ed.\ H.J.A.\ R\"ottgering, P.N.\ Best, \& 
M.D.\ Lehnert (Roy.\ Neth.\ Acad.\ Sci.: Amsterdam), 49
\bibitem[Veilleux \& Rupke(2002)]{vei02} Veilleux, S. \& Rupke, D. 2002, \apjl, 565, L63 
\bibitem[Villar-Mart{\'{\i}}n \etal(2003)]{vil03} Villar-Mart{\'{\i}}n, M., Vernet, J., di Serego Alighieri, S., Fosbury, R., Humphrey, A., \& Pentericci, L.\ 2003, \mnras, 346, 273 
\bibitem[Zylka(1998)]{zyl98} Zylka, R., The MOPSI Cookbook, IRAM

\end{thebibliography}
\end{document}